\documentclass[11pt,twoside]{article}

\usepackage{asp2004}
\usepackage{epsf}
\usepackage{psfig}
\usepackage{lscape}

\begin{document}
\title{Cosmic Infrared Background: Resolved, Unresolved
and {\it Spitzer} Contributions. }   
\author{Herv\'e Dole, Guilaine Lagache, Jean-Loup Puget}   
\affil{Institut d'Astrophysique Spatiale, b\^atiment 121, Universit\'e Paris Sud XI,
F-91405 Orsay Cedex, France}    

\begin{abstract} 
The Cosmic Infrared Background (CIB) peaks in the Far-Infrared (FIR),
and its Spectral Energy Distribution (SED) is now well
constrained. Thanks to recent facilities and {\it Spitzer}, the
populations contributing to the CIB are being characterized: the
dominant galaxy contributions to the FIR CIB are Luminous IR galaxies
(LIRGs) at $0.5 \le z \le 1.5$ and, to the submm CIB, Ultra-LIRGs at
$z \ge 2$.  These populations of galaxies experience very high rates
of evolution with redshift. Because of confusion, the CIB is (and will
remain in some domains) partially resolved and its contributing
galaxies SEDs are not well constrained. We discuss all these aspects
and show how confusion limits {\it Spitzer} observations, and how to
overcome it in order to study the unresolved part of the CIB.
\end{abstract}



\section{Introduction}
A Cosmic Background that would trace the peak of the star--formation
and metal production in galaxy assembly has long been predicted.  Its
discovery in 1996 by Puget et al. (see the review by Hauser \& Dwek
2001), together with recent cosmological surveys in the infrared (IR)
and submillimeter, has revolutionized our view on star formation at
high redshift. It has become clear that a population of galaxies
radiating most of their power in the far-IR (the so-called ``IR
galaxies'') contributes an important part of the whole galaxy build-up
in the Universe. Since 1996, detailed (and often painful)
investigations of the high-redshift IR galaxies have resulted in
spectacular results that are reviewed in Lagache et al. (2005).  In
this paper we detail briefly the sources of the Cosmic IR Background
(CIB, Sect. 2) and their average evolution properties (Sect. 3). We
discuss the extragalactic confusion noise, one of the most limiting
factor of Far-IR cosmological surveys (Sect. 4) and give first hints
on how we can overcome it with {\it Spitzer} and other facilities
(Sect. 5).

\section{Sources of the Cosmic IR Background: Redshift Distributions}
\begin{figure}[!ht]
\plotone{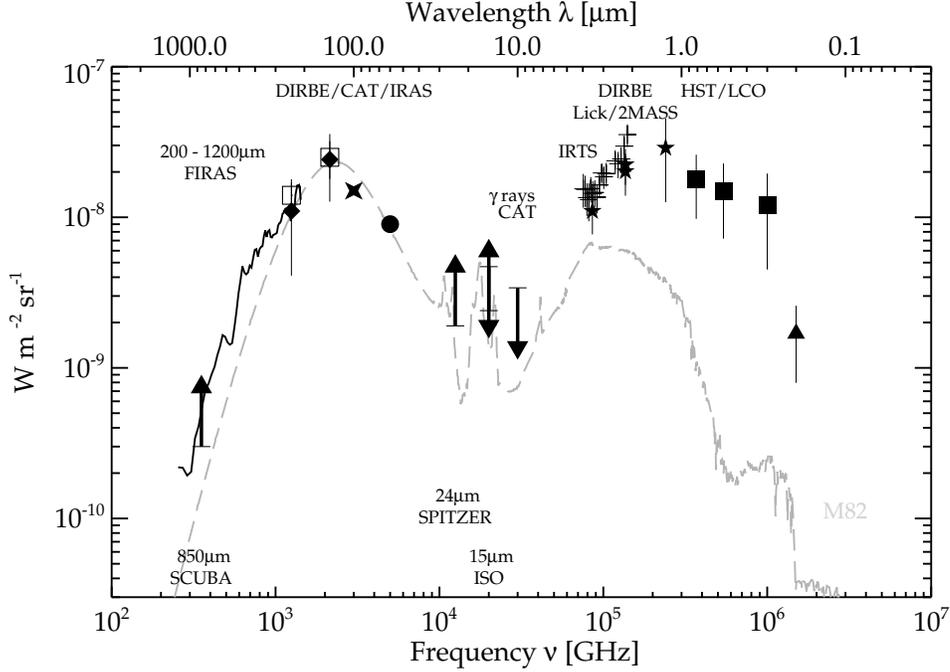}
\caption{The extragalactic background over 3 decades of frequencies from the
ultraviolet to millimeter wavelengths. Only strongly constraining
measurements have been reported. We show for comparison in gray an SED
of M82 (Chanial, 2003), a starburst galaxy at L=3 10$^{10}$
L$_{\odot}$, normalized to the peak of the CIB at 140~$\mu$m. CIB
measurements are, by decreasing frequency: Armand et al. (1994) at
2000~$\AA$; Bernstein et al. (2002) at 300, 555 and 814 nm using the
{\it HST} and Las Campanas Observatory; Mastumoto et al. (2004)
between 2.2 and 4~$\mu$m using the {\it IRTS}; Gorjian et al. (2000)
at 2.2 and 3.3~$\mu$m using {\it DIRBE} and {\it Lick}; Wright (2001)
at 1.25 and 2.2~$\mu$m using {\it DIRBE} and {\it 2MASS}; Renault et
al. (2001) upper limits at 10 and 15~$\mu$m using the {\it CAT} in the
$\gamma$-rays; Elbaz et al. (1999) lower limit at 15~$\mu$m using
galaxy counts with {\it ISOCAM}; Papovich et al. (2004) lower limit at
24~$\mu$m using galaxy counts with {\it Spitzer/MIPS}; An estimate of
the CIB at 60~$\mu$m from Miville-Desch\^enes et al. (2002) using {\it
IRAS}; Renault et al. (2001) at 100~$\mu$m using {\it CAT} and {\it
DIRBE}; Lagache et al. (2000) at 140 and 240~$\mu$m using {\it DIRBE}
and WHAM; Hauser et al. (1998) at 140 and 240~$\mu$m using {\it
DIRBE}; Smail et al. (2002) lower limit at 850~$\mu$m using galaxy
counts with {\it SCUBA}; Lagache et al. (2000) spectrum between
200~$\mu$m and 1.2mm using {\it FIRAS}.}
\label{fig:CIB_tot}
\end{figure}

The full Cosmic Extragalactic Background Spectral Energy Distribution
(SED) is shown in figure~\ref{fig:CIB_tot}. For clarity, we only
plotted the most recent and strongly constraining measurements. The
figure clearly shows that the optical and the IR cosmic backgrounds
are well separated.  The first surprising result is that the power of
the IR part (the CIB, Cosmic IR Background) is comparable to the power
of the optical one, although we know that locally the IR output of
galaxies is only one third of the optical (Soifer \& Neugebauer
1991). This implies a much stronger evolution of the IR luminosity of
IR galaxies than of the optical ones.  A second important property to
notice is the long wavelength ($\lambda \ge 200~\mu m$) behavior: the
CIB slope ($B_\nu \propto \nu^{1.4}$, Gispert et al. 2000), is much
shallower than the long wavelength spectrum of galaxies
(fig.~\ref{fig:CIB_tot}).  This implies that the millimeter CIB is not
due to the millimeter emission of the galaxies making the bulk of the
emission at the peak of the CIB (at $\sim 170~\mu m$). The
implications in terms of energy output have been drawn by e.g. Gispert
et al. (2000). The IR production rate per comoving unit volume 1)
evolves faster between redshift zero and 1 than the optical one and 2)
has to stay constant at higher redshifts up to redshift 3 at least.

\begin{figure}[!ht]
\plotone{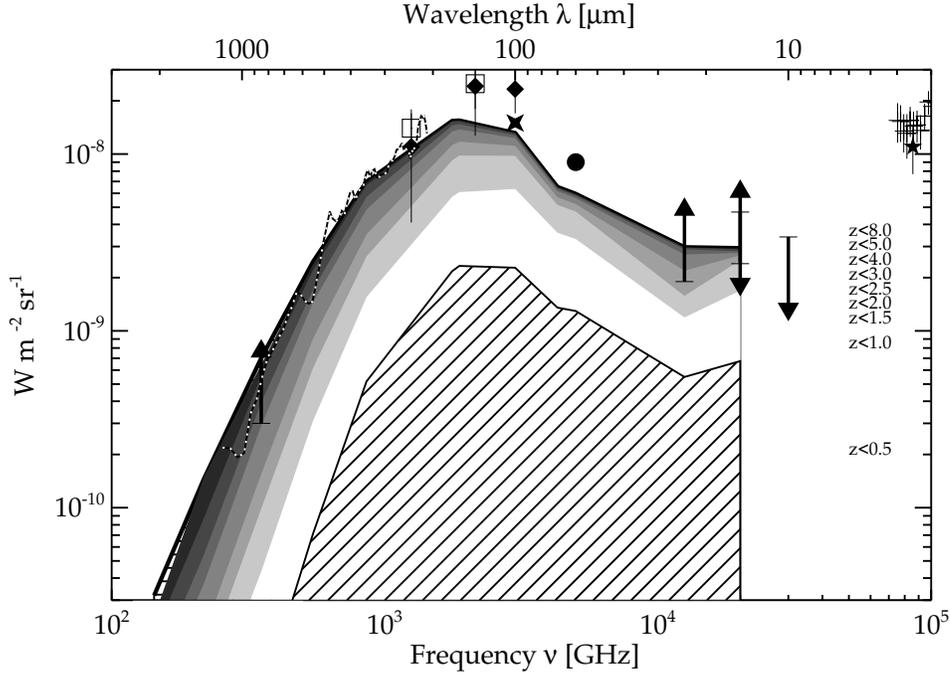}
\caption{Cumulative contribution by redshift (from 0.5 to 8) of
galaxies to the Cosmic IR Background, from the model of Lagache et
al. (2004). Measurements of the CIB are reported with the same symbols
as in figure~\ref{fig:CIB_tot}, except for the {\it FIRAS} spectrum,
represented in black and white dashes for visibility. z$<$0.5: oblique
shaded area; z$<$1.0: white area; z$<$1.5, 2.0, 2.5, 3.0, 4.0, 5.0:
darker gray areas; z$<$8: solid line and horizontal shaded
region. Notice the non negligible contribution of 5$<$z$<$8 galaxies
to the CIB, at wavelengths above 600~$\mu$m.  }
\label{fig:CIB_fraction_spec}
\end{figure}

In figure~\ref{fig:CIB_tot}, showing the CIB and a galaxy SED, we see
that contributions from galaxies at various redshifts are needed to
fill the CIB SED shape.  The bulk of the CIB in energy, i.e. the peak
at about 150~$\mu$m, is not yet resolved in individual sources, but
one dominant contribution to the CIB peak can be inferred from the
{\it ISOCAM} deep surveys. {\it ISOCAM} galaxies with a median
redshift of $\sim$0.7 resolve about 80\% of the CIB at
15~$\mu$m. Elbaz et al. (2002) separate the 15~$\mu$m galaxies into
different classes (ULIRGs, LIRGs, Starbursts, normal galaxies and
AGNs) and extrapolate the 15~$\mu$m fluxes to 140~$\mu$m using
template SEDs. A total brightness of (16 $\pm$ 5) nW m$^{-2}$
sr$^{-1}$ is found, that makes about two thirds of the CIB observed
value at 140~$\mu$m by {\it COBE/DIRBE}. Hence, the galaxies detected
by {\it ISOCAM} are responsible for a large fraction of energy of the
CIB. About one half of the 140~$\mu$m CIB is due to LIRGs and about
one third to ULIRGs. However, these {\it ISOCAM} galaxies make little
contribution to the CIB in the millimeter and submillimeter. At those
wavelengths, the CIB must be dominated by galaxies at rather high
redshift for which the SED peak is shifted. This effect and the
redshift contribution to the CIB are illustrated in
figure~\ref{fig:CIB_fraction_spec}.  We clearly see that the
submillimeter/millimeter CIB contains information on the total energy
output by high-redshift galaxies (z$>$2). This is supported by the
observed redshift distribution of the {\it SCUBA} sources at
850~$\mu$m that makes about 60\% of the CIB and have a median redshift
of 2.4 (Chapman et al. 2003).

\begin{table}[here]
\begin{center}
\begin{tabular}{|l|l|l|l|} \hline
Wavelength & 20\% & 50\% & 80\% \\ \hline
15~$\mu$m & 0.5 & 1.0 & 1.3 \\
24~$\mu$m & 0.5 & 1.3 & 2.0 \\
70~$\mu$m & 0.5 & 1.0 & 1.5 \\
100~$\mu$m & 0.7 & 1.0 & 1.7 \\
160~$\mu$m & 0.7 & 1.3 & 2.0 \\
350~$\mu$m & 1.0 & 2.0 & 3.0 \\
850~$\mu$m & 2.0 & 3.0 & 4.0 \\
1.4~mm & 2.5 & 3.5 & 4.5 \\
2.1~mm & 2.0 & 3.5 & 5.0 \\ \hline
\end{tabular}
\end{center}
\caption{Redshift at which the Cosmic IR Background is resolved at 
20, 50 or 80\%. Numbers have been derived using the model of Lagache
et al. (2004).\label{tab:cib_resolution}}
\end{table}

Fifty percents of the CIB are made by galaxies at redshift below 1 at
15 and 70~$\mu$m, below 1.3 at 24 and 160~$\mu$m and below 2, 3 and
3.5 at 350, 840 and 2000~$\mu$m (see also Table
\ref{tab:cib_resolution}).  It is clear that, from the far-IR to the
millimeter, the CIB at longer wavelength probes sources at higher
redshifts.

So far, the most constraining surveys (in terms of resolving the CIB)
are done at 15, 24 and 850~$\mu$m with {\it ISOCAM (ISO)}, {\it MIPS
(Spitzer)} and {\it SCUBA (JCMT)}. The deep field observations resolve
about 80, 70 and 60\% of the CIB, respectively (Elbaz et al., 2002;
Papovich et al., 2004; Smail et al., 2002). As shown above, these
surveys probe the CIB in well-defined and distinct redshift ranges,
with median redshift of 0.7 (Liang et al. 2004), $\sim$1.1 (L. Yan,
priv. comm.), and 2.4 (Chapman et al. 2003) at 15, 24 and 850~$\mu$m
respectively. Such well-defined redshift range selection can be
understood by looking at the K-corrections (defined as $K(L,z) =
\frac{L_{\nu(1+z)}}{L_{\nu(z=0)}}$ where $L_{\nu(z=0)}$ is the
rest-frame luminosity).  This correction is specific of the spectrum
of the population considered at a given luminosity and
redshift. Taking the standard SEDs of local normal starbursting
galaxies (i.e. with PAHs features in the mid-IR), we observe in the
K-correction at 15~$\mu$m a hump at z$\sim$1 associated with the
coincidence of the 6 to 9~$\mu$m aromatic features and the {\it
ISOCAM} filter, and at 24~$\mu$m a hump at the same redshift for the
24~$\mu$m {\it MIPS} filter associated with the 11-14~$\mu$m set of
aromatic features.  Furthermore, a second hump is expected for the
{\it MIPS} 24~$\mu$m filter at z$\sim$2, which corresponds to the
redshifted 6 to 9~$\mu$m features centered on this{\it MIPS} filter
(Dole et al., 2003; Lagache et al., 2004). There is still a debate
concerning the presence of this second hump in the 24~$\mu$m sources
redshift distribution observations (e.g. P\'erez-Gonz\'alez et
al. 2005 vs L. Yan \& P. Choi, {\it this volume}), even if growing
evidences show the existence of PAHs at large redshift (e.g. Elbaz et
al., 2005).  At longer wavelengths, in the submillimeter and
millimeter, the negative K-correction becomes very effective leading
to an almost constant observed flux for galaxies of the same total IR
luminosity between redshifts 1 and 5.

\section{Infrared Galaxy Cosmic Evolution}
\begin{figure}[!ht]
\plotfiddle{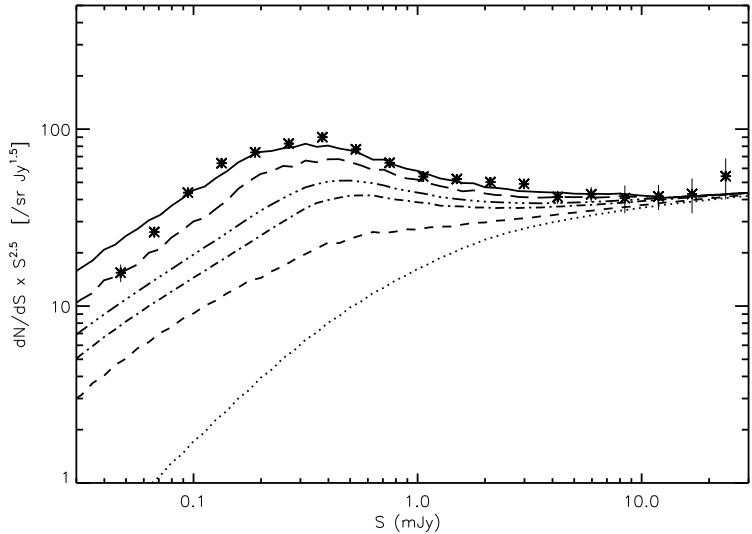}{7.0cm}{0}{60}{60}{-150}{0}
\caption{Redshift contribution to the number counts at 24~$\mu$m from
the model of Lagache et al. (2004, their figure~7). The dot, dash,
dash-dot, dash-3 dot, long-dash correspond to the number counts up to
redshifts 0.3, 0.8, 1, 1.3 and 2 respectively. Data are from Papovich
et al. (2004). }
\label{fig:24um_counts}
\end{figure}

A remarkable property of the IR galaxies is their extremely high rates
of evolution with redshift, exceeding those measured for galaxies at
other wavelengths, and comparable or larger than the evolution rates
observed for quasars. Number counts at 15 and 24~$\mu$m show a
prominent bump peaking at about 0.4~mJy and 0.3~mJy respectively.  At
the peak of the bump, the counts are more than one order of magnitude
above the non-evolution models. At 15~$\mu$m, data require a
combination of a (1+z)$^3$ luminosity evolution and (1+z)$^3$ density
evolution for the starburst component at redshift lower than 0.9 to
fit the strong evolution. While it has not been possible with {\it
ISOCAM} to probe in detail the evolution of the IR luminosity
function, {\it Spitzer} data at 24~$\mu$m give for the first time
tight constraints up to redshift 1.2 (Le Floc'h et al. 2005). A strong
evolution is noticeable and requires a shift of the characteristic
luminosity L$^{\star}$ by a factor (1+z)$^{4.0\pm0.5}$. Le Floc'h et
al. (2005) find that the LIRGs and ULIRGs become the dominant
population contributing to the comoving IR energy density beyond
z$\sim$0.5-0.6 and represent 70$\%$ of the star-forming activity at
z$\sim$1. Those findings are in good agreement with the predictions
from Lagache et al. (2004). This model constrains in a simple way the
IR luminosity function evolution with redshift, and fits all the
existing source counts consistent with the redshift distribution, the
CIB intensity, and, for the first time, the CIB fluctuation
observations from the mid-IR to the submillimeter range.  In this
model, we assume that IR galaxies are mostly powered by star formation
and hence we use SEDs typical of star-forming galaxies Although some
of the galaxies will have AGN-dominated SEDs, they are a small enough
fraction that they do not affect the results significantly.  We
therefore construct 'normal' and starburst galaxy template SEDs: a
single form of SED is associated with each activity type and
luminosity.  We assume that the Luminosity Function (LF) is
represented by these two activity types and that they evolve
independently and we search for the form of evolution that best
reproduces the existing data.  An example of two cosmological
implications of this model is that (1) the PAHs features remain
prominent in the redshift band 0.5 to 2.5 and (2) the IR energy output
has to be dominated by $\sim$3 10$^{\rm 11}$~L$_{\rm \odot}$ to
$\sim$3 10$^{\rm 12}$~L$_{\rm \odot}$ galaxies from redshift 0.5 to
2.5.  The excellent agreement between the model and all the available
observational constraints makes this model a likely good
representation of the average luminosity function as a function of
redshift and a useful tool to discuss observations and models
(e.g. figure~\ref{fig:24um_counts} showing the contribution by
redshift to the MIPS 24~$\mu$m source counts). Its rather simple
assumptions like the single parameter sequence of SEDs for starburst
galaxies is certainly not accounting for some of the detailed recent
observations (e.g. Lewis et al. 2005) but probably do not affect
seriously the redshift evolution of the averaged properties which are
what is modeled.\\

The comoving luminosity density produced by luminous IR galaxies was
more than 10 times larger at z$\sim$1 than in the local Universe. For
comparison, the B-band luminosity density was only 3 times larger at
z=1 than today. Such a large number density of LIRGs in the distant
Universe could be caused by episodic and violent star formation
events, superimposed to relatively small levels of star formation
activity (Hammer et al. 2005).  These events can be associated to
major changes in the galaxy morphologies. \\

\begin{figure}[!ht]
\plotone{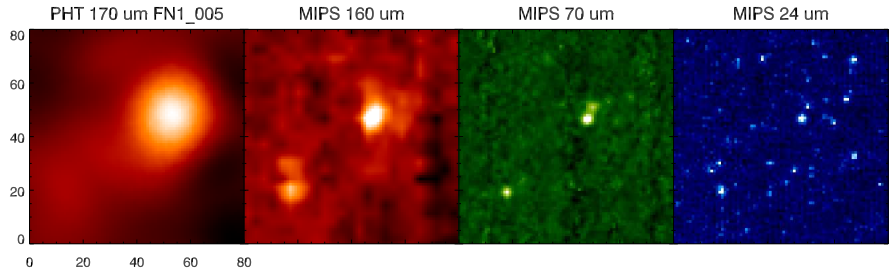}
\caption{Effects of confusion in the far-IR. Observation of a
source in the ELAIS-N1/FIRBACK field, in a 400x400 square arcsecond
box. All plates have been resampled to 5 arcseconds per pixel, which
oversamples the far IR maps but undersamples the mid-IR map.  From
left to right: 373 mJy {\it ISOPHOT} 170~$\mu$m source with about 128s
of integration (FIRBACK survey, Dole et al., 2001); Labels indicate
the 5 arcsecond pixels; {\it Spitzer} {\it MIPS} 160~$\mu$m with about
16s of integration (SWIRE survey, Lonsdale et al., 2004); {\it MIPS}
70~$\mu$m with about 80s of integration (SWIRE); {\it MIPS} 24~$\mu$m
with about 160s of integration (SWIRE).  Notice (1) the {\it ISO}
170~$\mu$m source is marginally resolved with {\it MIPS} 160, and is
unambiguously resolved at 70~$\mu$m and 24~$\mu$m; (2) the two fainter
{\it MIPS} 160~$\mu$m resolved sources (bottom left) create
fluctuations in the {\it ISO} 170~$\mu$m map that produce the
confusion noise when the resolution is limited.  }
\label{fig:confusion}
\end{figure}

The large fraction of the background resolved at 850~$\mu$m has
interesting consequences. It shows very directly that if the sources
are at redshift larger than 1 (as confirmed by the redshift surveys),
the IR luminosity of the sources dominating the background is larger
than $10^{12} L_{\odot}$, in agreement with the predictions of Lagache
et al. (2004), thus making a population with a very different IR
luminosity function than the local or even the z=1 luminosity
function. The link between this population at high z and what has been
seen around redshift one (almost by {\it ISO}) will be done by {\it
Spitzer}/{\it MIPS} observations at 24~$\mu$m.  {\it ISOCAM} galaxies
contribute to about 2/3 of the energy peak of the CIB (Elbaz et
al. 2002).  The remaining fraction is likely to be made of sources in
the redshift range 1.5 to 2.5. The presently detected submillimeter
galaxies with luminosity $10^{12} L_{\odot}$ have an almost constant
flux between redshift 1.7 and redshift 2.5 at 24~$\mu$m (similar to
the constant flux at 850~$\mu$m between redshift 1 and 5). The {\it
MIPS} 24~$\mu$m deep surveys (e.g. Papovich et al. 2004) reach an 80\%
completeness of 80~$\mu Jy$ (and many sources are fainter) and thus
can detect all these galaxies when they are starburst-dominated.
Considering the efficiency of {\it MIPS} to cover large areas of the
sky with a good sensitivity, it is likely that 24~$\mu$m surveys will
become the most effective way to search for luminous starburst
galaxies up to z= 2.5 and up to 3 for the most luminous ones.  In the
near future when a proper census of ULIRGs up to z~$\simeq$~3 will be
done, the fraction of the CIB at $\sim$1 mm not accounted for should
give an indication of the contribution from sources at larger
redshifts.  Deep surveys around 1-2 mm are the only obvious tool to
get most of these sources. But the limiting factors of the surveys are
not only detector sensitivity or photon noise, there is also
confusion.

\section{Confusion in the Mid- and Far-Infrared}
Even if the term ``confusion'' might be vague, we can nevertheless
define the confusion in general as the degradation of a high spatial
frequency signal due to a poor angular resolution (see
fig.~\ref{fig:confusion}). By degradation, we mean the measurement of
a source is affected either by a poorer quality of photometry, or a
lower detectability. By high frequency signal we mean small-scale
observed structures compared to the beam, usually extragalactic
point-sources. Finally, by poor angular resolution, we mean a large
beam or a poor point spread function sampling compared to the studied
structure, usually a set of point sources.

Predicting or measuring confusion depends on the scientific goal of
the measurement (e.g. Helou \& Beichman, 1990; Dole et al., 2003;
Lagache et al., 2003): performing an unbiased far-IR or submillimeter
survey and getting a complete sample has different requirements than
following-up in the far-IR an already known near-IR source to get a
SED and/or a photometric redshift. In the former case, one has to
tightly control the statistical properties of the whole sample; in the
latter case, completeness is irrelevant, and an even low photometric
accuracy is adequate. We thus favor the use of a qualifier, like
``unbiased confusion'' for the former case. New techniques are being
developed to use an {\it a priori} information at shorter wavelength
(e.g. 8~$\mu$m with {\it Spitzer}/IRAC and 24~$\mu$m with {\it MIPS},
respectively) to infer some statistical properties (like source
density or SED) of sources at longer wavelength (e.g. 24 or
160~$\mu$m, respectively), and thus to beat the unbiased confusion.

Predicting the unbiased confusion (for instance Condon, 1974;
Franceschini et al., 1989; Helou \& Beichman, 1990; Rieke et al. 1995;
Dole et al., 2003; Lagache et al., 2003; Takeuchi \& Ishii, 2004;
Negrello et al., 2004) requires the knowledge of at least the number
counts distribution of the galaxies. In practice, models (validated at
some point by observations) are used. Since the slope of the counts in
a $Log(N) - Log(S_{\nu})$ diagram is varying with the flux density
$S_{\nu}$, the fluctuation level of faint sources below $S_{\nu}$ will
also vary; this fluctuation level gives an estimate of the unbiased
confusion using a photometric criterion (Dole et al., 2003; Lagache et
al., 2003). At very faint fluxes, when the background is almost
resolved, the photometric criterion will obviously give a very small
value for the unbiased confusion level, but the observations will be
limited by the confusion due to the high density of faint resolved
sources. Thus, another criterion, the source density criterion for
unbiased confusion (SDC, Dole et al., 2003, 2004), needs to be
computed and compared to the photometric criterion.  In the IR and
submillimeter range below 300~$\mu$m, the unbiased confusion is in
general better predicted by the source density criterion for current
and future facilities, since the angular resolution has improved
(e.g. from ISO to {\it Spitzer}). At longer wavelengths, the
photometric criterion still dominates. \\

\begin{figure}[!ht]
\plotfiddle{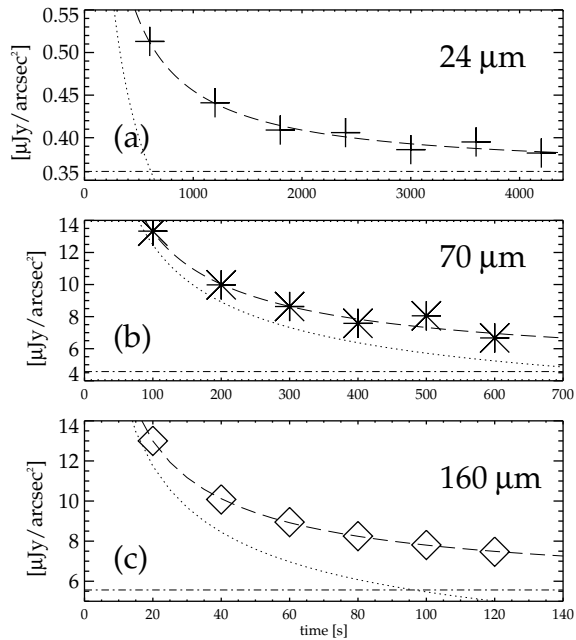}{8.5cm}{0}{50}{50}{-125}{0}
\caption{Evolution $\sigma_{tot}$ (resulting contribution
from the confusion noise and instrument noise) as a function of
integration time, with a fit (dash) of the form: $\sigma_{tot}^2 =
\sigma_{inst}^2 + \sigma_{conf}^2 = At^{-1} + C^2$.  dot-dash:
constant term $C$. dot: $\sqrt{A/t}$ term.  Top panel (a): 24~$\mu$m.
Middle panel (b): 70~$\mu$m. Bottom panel (c): 160~$\mu$m.  Notice the
different scales in time (seconds) and $\sigma_{tot}$
($\mu$Jy/arcsec$^2$). From Dole et al., 2004b.  }
\label{fig:confusion_sigma}
\end{figure}

In Dole et al. (2004b), we derived the confusion limits for MIPS.
Using data at 24, 70 and 160~$\mu$m, the source density measured by
Papovich et al. (2004) and Dole et al. (2004a) together with the
modeling of Lagache et al. (2004) has allowed us to derive the
confusion limits for {\it Spitzer} in the mid to far IR.  We tested
the model results with a Monte Carlo simulation at 24~$\mu$m and with
a fluctuations analysis at all 3 wavelengths
(fig.~\ref{fig:confusion_sigma}).  The agreement is uniformly very
good.  We determined the 5-$\sigma$ unbiased confusion limits due to
extragalactic sources: $56 \mu$Jy, 3.2 and 40 mJy at 24, 70 and
160~$\mu$m, respectively. At 24 and 70~$\mu$m, confusion is mostly due
to the high density of resolved sources, and at 160~$\mu$m, confusion
is mainly due to faint unresolved sources. Studying the FIR
fluctuations at this wavelength is thus a tool to constrain the nature
of the faint galaxies, beyond the confusion limit.

What can be done with these numbers ? Generally speaking, you should
stop integrating down to these levels, unless you know the existence
of an object (e.g. from shorter wavelength observations) you want
extract photometry from. It might thus be useful to integrate deeper
than these limits, but the analysis will require some care. For
instance, a sample of detected extragalactic faint sources will likely
be highly incomplete; this might be a problem for statistical studies,
but this can be ignored (or has to be appropriately quantified) if one
wants photometry of known objects.

These numbers were derived both from the data and a model, and they
explicitly made the assumption of a Poisson source distribution.
Source clustering might in principle bias these estimates. Takeuchi \&
Ishii (2004) derived analytic formulas taking into account the source
clustering, and showed that this effect is at the $\sim 10\%$ level.\\

Confusion is also created in the IR by the presence of Galactic cirrus
(Low et al., 1984), even if its complex structure has little power at
small scales (Gautier et al., 1992; Herbstmeier et al., 1998;
Miville-Desch\^enes et al., 2003, Kiss et al., 2001 \& 2003; Ingalls
et al., 2004). Usually, IR cosmological surveys are performed in high
Galactic latitude cirrus-free regions, with a column-density as low as
possible. For instance, only 2\% of the sky have $N_{HI} \le 1.0
\times 10^{20} \, cm^{-2}$. Table~\ref{tab:cirrus_sky} gives more
details about the fraction of the sky cleaner than certain
column-densities.  A common problem arises when cosmological fields,
chosen in the visible range, happen to fall in high Galactic cirrus
contamination fields, avoiding any efficient IR follow-up, like the
22h VIMOS field 2217+0024 with a 100~$\mu$m brightness of about 4
MJy/sr.\\

\begin{table}[here]
\begin{center}
\begin{tabular}{|l|c|} \hline
N(HI)/$ cm^{-2}$ & Sky Fraction \\ \hline
$\le 1.0 \times 10^{20}$ & 2\% \\
$\le 1.25 \times 10^{20}$ & 5\% \\
$\le 1.6 \times 10^{20}$ & 10\% \\
$\le 2.0 \times 10^{20}$ & 17\% \\
$\le 2.2 \times 10^{20}$ & 20\% \\
$\le 3.0 \times 10^{20}$ & 30\% \\
$\le 3.8 \times 10^{20}$ & 40\% \\
$\le 5.0 \times 10^{20}$ & 52\% \\ \hline
\end{tabular}
\end{center}
\caption{Fraction of the sky with HI column-densities lower than
certain values. These numbers are derived from the HI Burton \&
Hartman (1994) Leiden/Dwingeloo survey, assuming an all sky coverage.
\label{tab:cirrus_sky}}
\end{table}

A number of cryogenically-cooled space telescopes have been proposed
for the MIR, the FIR and the submillimeter spectral ranges. Table~2 of
Dole et al. (2004b) summarizes the main characteristics of some of
these observatories.  Herschel (Pilbratt, 2001), JWST (Gardner, 2003),
SPICA (Matsumoto, 2003) and SAFIR (Yorke 2002), have at least one
photometric channel in common with MIPS.  As examples, Dole et
al. (2004b) focused on Herschel-PACS at 75 and 170~$\mu$m, on
JWST-MIRI at 24~$\mu$m, and on SPICA and SAFIR at 24, 70 and
160~$\mu$m, assuming in each case that the MIPS filters will be
used. In Table~3 of Dole et al. (2004b), we use the confusion level
given by the SDC, and compute the fraction of the CIB potentially
resolved into sources.  In the MIR, a significant step will be made
with the 4m-class space telescope: as an example, SPICA would
potentially resolve 98\% of the CIB at 24~$\mu$m. All ($>99$\%) of the
CIB would be resolved with JWST or SAFIR (although doing so with JWST
would require extremely long integrations).  In the FIR, Herschel
would resolve a significant fraction of the CIB at 70 and 160~$\mu$m
(resp. 93 and 58\%, again with extremely long integrations). SAFIR
will ultimately nearly resolve all of it ($>94$\%).

Predictions of the unbiased confusion, for various telescope diameters
and wavelengths, will be soon available on our
website\footnote{http://lully.as.arizona.edu/Model}.

\section{Beating the Unbiased Confusion}
Observing in the FIR and submm spectral ranges is relevant to
characterize the galaxies responsible of the CIB at redshifts $z \ge
1.3$ (Table~\ref{tab:cib_resolution}) but the unbiased confusion
usually avoids to directly probe these galaxies. What can be done ? As
already mentioned, one method consists in extrapolating the FIR and
submm properties of the galaxies from the MIR (or from the radio)
spectral properties, based on controlled samples having FIR and submm
data. These controlled samples are usually small (e.g. Appleton et al.,
2004). This has the advantage of allowing to apply this extrapolation
to large samples like the new IRAC or MIPS catalogs from GTO or legacy
surveys, but the disadvantage of being dependent on the extrapolation
based on a small sample, and of not allowing any object by object
study. More observations are thus required in the FIR and submm to
extend these controlled samples; more {\it Spitzer} GO programs should
address this need in the near future.

\begin{figure}[!ht]
\plotone{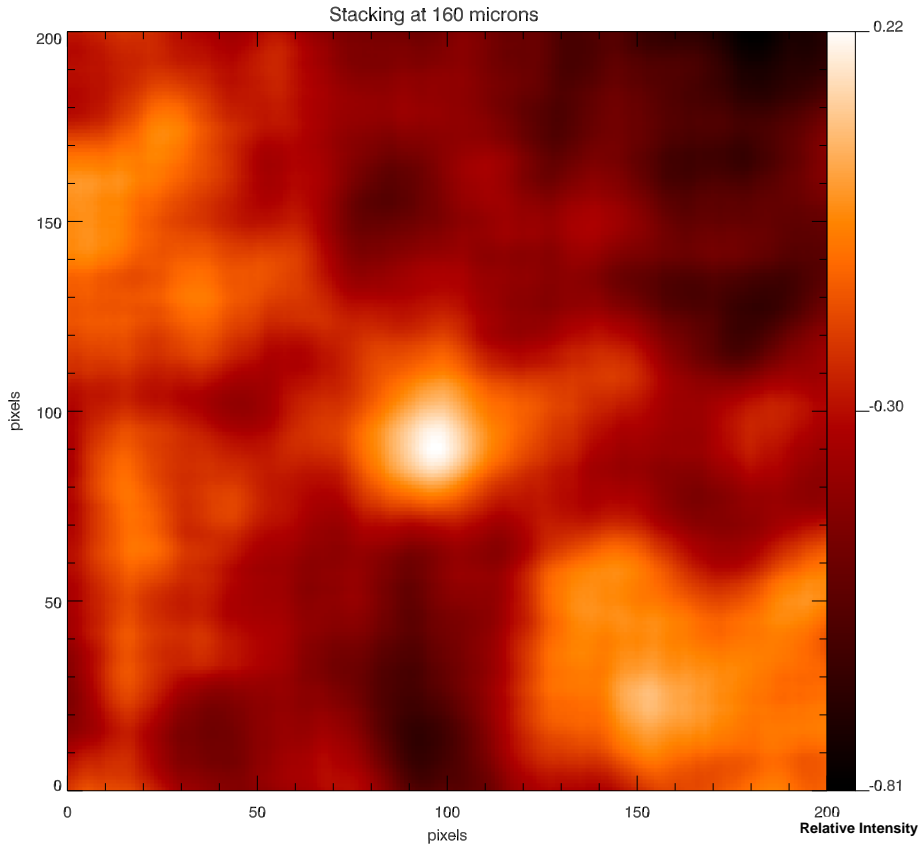}
\caption{Example of a stacking analysis at 160~$\mu$m. 24~$\mu$m
sources in the CDFS are selected to have flux densities $80 \le S_{24}
\le 83 \, \mu$Jy (330 sources). Images at 160~$\mu$m are stacked at
the positions of the 24~$\mu$m sources. There is a clear detection in
the resulting stacked 160~$\mu$m image in the center, with the
expected FWHM. Pixel scale here is 1.25 arcseconds/pixel.}
\label{fig:stack_160}
\end{figure}

Another method consists in characterizing already known submm (or FIR)
sources. Egami et al. (2004) studied VLA and SCUBA sources in the NIR
and MIR, and Ivison et al. (2004) did the same of MAMBO 1.2mm sources
and both showed that IRAC and MIPS quickly detect high redshift ($0.5
\le z \le 3.5$) massive galaxies, dominated by star formation. 
Although being successful, this method rely also on small samples.

Finally, a powerful method consists in making biased observations:
using shorter wavelength data (like IRAC 8.0~$\mu$m or MIPS 24~$\mu$m)
or high angular resolution data (e.g. VLA 1.4~GHz) to select a
physically homogeneous sample, one can extract photometry in the FIR
or submm. If the targets fluxes are below the MIPS FIR or SCUBA submm
detection limits, one can make use of a stacking analysis to increase
the S/N of FIR or submm data.  Frayer et al. (2004) used this
technique to extract color properties of SCUBA galaxies at
70~$\mu$m. Also, Sergeant et al. (2004) stacked 450 and 850~$\mu$m
images at the positions of IRAC 5.8 \& 8~$\mu$m sources, and detected
a positive deviation, allowing them to give a constrain on the
contribution of IRAC sources to the submm CIB.

We present here an example of such stacking analysis at 160~$\mu$m for
illustration. We first selected a sample of 24~$\mu$m galaxies with
flux densities $80 \le S_{24} \le 83 \, \mu$Jy in the CDFS, and found
330 sources (Papovich et al., 2004).  We then stacked the
corresponding 160~$\mu$m data. Figure~\ref{fig:stack_160} shows the
result of a clear detection. This illustrates the power of stacking
analysis to probe fainter galaxy population using an {\it a priori}
information from shorter wavelengths.

Not only we can probe individual galaxies deeper into the confusion in
the MIR using an {\it a priori} information at shorter wavelength
(e.g. NIR), but we can also probe galaxies statistically in the
FIR. In this case, obviously a careful sample selection is required to
interpret the resulting photometry.

\section{Conclusion}
The Cosmic Infrared Background, peaking in the FIR, is made of
contributions by galaxies located at different redshifts. The FIR CIB
has most contribution from $0.5 \le z \le 1.5$ LIRGs, and the submm
has most contributions from $z \ge 2$ ULIRGs. The FIR properties of
these IR bright galaxies are not known in detail, mainly because of
the unbiased confusion in this spectral range. New {\it Spitzer}
results are starting to refine our knowledge of these galaxies.
Beating the unbiased confusion is one of the challenges of the
analysis of today's cosmological surveys, and early results show that
many methods are promising and successful.

Furthermore, the efficiency of MIPS 24~$\mu$m to cover large sky areas
to a great depth is a unique tool to search high redshift ULIRGs
responsible of the submm CIB. There might be another {\it Spitzer}
legacy consisting to cover wider areas to pinpoint this population,
which might be characterized using also submm data from Herschel (and
Planck), either on an individual basis or statistically.

\acknowledgements 
We wish to thank the organizers. Special thanks to the MIPS GTO team
for a fruitful and enjoyable collaboration. Thanks to the FLS, SWIRE
and GOODS teams, as well as IRAC, IRS and SSC people, for the great
work providing the community with outstanding {\it Spitzer} data.


\end{document}